\documentclass[aps,showpacs,prl,twocolumn,preprintnumbers,amsmath,
amssymb]{revtex4}
\usepackage{graphicx}
\usepackage{dcolumn}
\usepackage{bm}
\def\Sc#1{\textsc{#1}}

\begin{document}

\preprint{Published in {\it Int. J. Mod. Phys.} {\bf B}~19, 775
(2005) [Preprint number: ITP-UU-2004/04]}

\title{Composite fermions close to the one-half filling of the lowest
Landau level revisited}

\author{Behnam Farid}
\email{B.Farid@phys.uu.nl}
\affiliation{Spinoza Institute and Institute for Theoretical Physics,
Department of Physics and Astronomy,
University of Utrecht, \\
Leuvenlaan 4, 3584 CE Utrecht, The Netherlands \\
and
Cavendish Laboratory, Department of Physics,
University of Cambridge,\\
Madingley Road, Cambridge CB3 0HE, United Kingdom}

\date{October 11, 2004}

\begin{abstract}
By strictly adhering to the microscopic theory of composite
fermions (CFs) for the Landau-level filling fractions $\nu_{\rm e}
= p/(2 p + 1)$, we reproduce, with remarkable accuracy, the
surface-acoustic-wave (SAW)-based experimental results by Willett
and co-workers concerning two-dimensional electron systems with
$\nu_{\rm e}$ close to $\frac{1}{2}$. Our results imply that the
{\sl electron} band mass $m_{\rm b}$, as distinct from the CF mass
$m_{\star}$, must undergo a substantial increase under the
conditions corresponding to $\nu_{\rm e} \approx \frac{1}{2}$. In
view of the relatively low aerial electronic densities $n_{\rm e}$
to which the underlying SAW experiments correspond, our finding
conforms with the experimental results by Shashkin {\it et al.}
[Phys. Rev. B {\bf 66}, 073303 (2002)], concerning two-dimensional
electrons in silicon, that signal sharp increase in $m_{\rm b}$
for $n_{\rm e}$ decreasing below approximately $2\times
10^{11}$cm$^{-2}$. We further establish that a finite mean-free
path $\ell_0$ is essential for the {\sl observed} linearity of the
longitudinal conductivity $\sigma_{xx}(q)$ as deduced from the SAW
velocity shifts.
\end{abstract}


\maketitle

{\it 1.~Introduction.---} The fractional quantum Hall effect
(FQHE) owes its existence to the electron-electron (e-e)
interaction. The fermionic Chern-Simons field theory in $2+1$
dimensions unifies the FQHE with the integer QHE (IQHE) whose {\sl
existence} does not depend on e-e interaction
\cite{Dai,Lopez,Halperin,Murthy}. This is effected through the
binding, brought about by the mediation of the Chern-Simons
action, of $2 n$, $n=1,2,\dots$, magnetic flux quanta to
electrons, whereby the composite particles, that is CFs
\cite{Jain}, are exposed to an effective magnetic flux density
$\Delta B$ whose corresponding IQH state determines the FQH state
for the electrons; the FQH state associated with $\nu_{\rm e} =
p/(2 n p + 1)$ is the IQH state of CFs in which $p$ lowest Landau
levels (LLs) are fully occupied. In this paper we consider $n=1$
for which the sequence $\nu_{\rm e} = p/(2 p + 1)$ and its
particle-hole conjugate $\nu_{\rm e}' {:=} 1 - p/(2 p + 1)$
approach $\nu_{\rm e} = \frac{1}{2}$ for $p\to \infty$. With
$\Delta B \equiv h n_{\rm e}/(e p)$, where $-e < 0$ is the
electron charge, these states thus correspond to small
$\vert\Delta B\vert$. The state corresponding to $\nu_{\rm
e}=\frac{1}{2}$ was proposed by Halperin, Lee and Read
\cite{Halperin} to be a compressible state of degenerate fermions
which, in the case of full polarization of the electron spins, is
characterised by a circular Fermi surface of radius $k_{\Sc
f}^{\Sc c\Sc f} = \sqrt{4\pi n_{\rm e}}$. This has been borne out
by several experiments
\cite{Willett1,Willett2,Kang,Goldman,Smet1}.

In this paper, which in all essential respects coincides with our
earlier unpublished work \cite{Farid0}, we particularly
concentrate on a series of experimental results by Willett and
co-workers \cite{Willett1,Willett2,Willett3}, concerning
two-dimensional electron systems (2DESs) with $\nu_{\rm e}$ at and
close to $\frac{1}{2}$ and establish that these can be remarkably
accurately reproduced within the framework of the microscopic
Chern-Simons field theory. To achieve this, it turns out to be
essential that the band-electron mass $m_{\rm b}$ is by one order
of magnitude greater than the customarily-assumed value: for GaAs
heterostructures, in which the 2DESs under consideration were
realised, $m_{\rm b}$ is customarily taken to be $0.067\times
m_{\rm e}$, where $m_{\rm e}$ is the electron mass in vacuum. In
what follows, we use the notation $m_{\rm b}^0 = 0.067 \times
m_{\rm e}$ and denote the electron band mass, as required for
reproducing the above-indicated experimental results, by $m_{\rm
b}$. Denoting the CF mass by $m_{\star}$, our numerical results
imply $m_{\rm b} \gtrsim 16 \times m_{\rm b}^0$ and $m_{\star}
\approx 0.5 \times m_{\rm b} \gtrsim 8 \times m_{\rm b}^0 \approx
0.54 \times m_{\rm e}$. This value is in good accord with the CF
mass as deduced both from the values of the energy gaps
$\Delta_{\nu_{\rm e}}$ separating the CF LLs
\cite{Du1,Du2,Manoharan,Willett3} and the amplitude of the
oscillations in $\varrho_{xx}$ for varying $\Delta B$
\cite{Leadly,Du3,Coleridge,Manoharan,Willett3}. Our finding with
regard to the increase of $m_{\rm b}$ in comparison with $m_{\rm
b}^0$ finds support in the independent experimental results by
Shashkin {\sl et al.} \cite{Shashkin} according to which $m_{\rm
b}$ (corresponding to two-dimensional electrons in silicon)
undergoes a sharp increase for $n_{\rm e}$ decreasing below
approximately $2 \times 10^{11}$ cm$^{-2}$ (see Fig.~3 in
\cite{Shashkin}), corresponding to the Wigner-Seitz radius $r_{\rm
s}$ greater than approximately $6$; that a critical value for
$n_{\rm e}$ associated with the divergence of $m_{\rm b}$ is not
specific to silicon is corroborated by recent theoretical
considerations concerning uniform states of idealised 2DESs, in
the absence of external magnetic field
\cite{MorawetzZhang1Asgari}. We expect that the optically-measured
CF mass enhancements (in comparison with $m_{\star} \approx 0.54
\times m_{\rm e}$), as reported by Kukushkin {\sl et al.}
\cite{Kukushkin}, are also ascribable to the increase of $m_{\rm
b}$ with respect to $m_{\rm b}^0$ \cite{Farid0}.

{\it 2.~Surface-Acoustic-Wave (SAW) experiments and theoretical
details.---} In the experiments by Willett {\sl et al.}
\cite{Willett1,Willett2,Willett3}, the relative change $\Delta
v_{\rm s}/v_{\rm s}$ in the velocity $v_{\rm s}$ of SAWs as well
as the damping $\kappa$ in their amplitudes, while propagating on
the surface of samples at distance $d$ from the 2DES, were
measured; here the change $\Delta v_{\rm s}$ is relative to the
$v_{\rm s}$ corresponding to the case where the conductivity of
the 2DES is infinitely large. Theoretically, $\Delta v_{\rm
s}/v_{\rm s}$ is determined by the `on-the-mass-shell' value of
the longitudinal conductivity $\sigma_{xx}(q,\omega)$ of the 2DES
as follows
\begin{equation}
\label{e1}
\frac{\Delta v_{\rm s}}{v_{\rm s}} = \frac{\alpha^2/2}{1 +
[\sigma_{xx}(q)/\sigma_{\rm m}]^2},
\end{equation}
where $\sigma_{xx}(q) \equiv \sigma_{xx}(q,\omega=v_{\rm s} q)$,
$\sigma_{\rm m}$ is a constant to be specified below, and
\cite{Simon1} $\alpha^2/2 =$
$(e_{14}^2\tilde\epsilon_{\rm r}/\{H \epsilon_0
\epsilon_{\rm r}^2\})$ $\vert F(q d) \vert^2$; here $e_{14} \approx
0.145$~C~m$^{-2}$ is the piezoelectric constant for Al$_x$Ga$_{1-x}$As,
$\tilde\epsilon_{\rm r} = \epsilon_{\rm r} \big(1 + \{(\epsilon_{\rm r}
- 1)/(\epsilon_{\rm r} + 1)\} \exp(-2 q d)\big)^{-1}$ ({\sl c.f.}
Eq.~(A4) in Ref.~\cite{Simon1}), $H\approx 28.8\times 10^{10}$~Nm$^{-2}$,
$\epsilon_0 = 8.854\dots\times 10^{-12}$~Fm$^{-1}$ denotes the vacuum
permittivity, $\epsilon_{\rm r}$ is the relative dielectric constant of
the {\sl bulk} of the host material which we take to be equal to $12.4$,
and $F(x) {:=} A_1 \exp(-a x) \sin(b x + c) + A_2 \exp(-x)$ ({\sl c.f.}
Eq.~(51) in Ref.~\cite{Simon1}), where $A_1 \approx 3.18$, $a \approx
0.501$, $b \approx 0.472$, $c\approx 2.41$ and $A_2 \approx -3.10$;
further,
\begin{equation}
\label{e2}
\sigma_{\rm m} {:=} \left. \frac{\omega e^2}{q^2 v(q)}\right|_{\omega
= v_{\rm s} q} \equiv 2 \epsilon_0 \tilde\epsilon_{\rm r} v_{\rm s},
\end{equation}
where in the second expression on the right-hand side we have replaced
the e-e interaction function $v(q)$ by the Coulomb function $v(q) =
e^2/(2 \epsilon_0\tilde\epsilon_{\rm r} q)$; we have further made use
of the dispersion of acoustic phonons and employed $\omega= v_{\rm s} q$,
where $v_{\rm s}$ stands for the sound velocity which in GaAs amounts
to $3010$~ms$^{-1}$.

Our calculations are based on \cite{Halperin} (see also Eq.~(208)
in Ref.~\cite{Simon2}) $\Delta v_{\rm s}/v_{\rm s} + i \kappa/q =
(\alpha^2/2) \big( \gamma - v(q) K_{00}(q,\omega= v_{\rm s}
q)\big)$, where $-K_{00}(q,\omega)$ describes the change in
$n_{\rm e}$ to linear order in the {\sl external} potential
\cite{Dai,Lopez,Halperin,Simon3}; the minus sign here has its
origin in our convention with regard to the sign of $e$. In the
literature, the constant $\gamma$ is invariably identified with
unity, this on account of the fact that $\Delta v_{\rm s}$ is the
deviation of the measured $v_{\rm s}$ with respect to the $v_{\rm
s}$ pertaining to the case where $\sigma_{xx}(q) \to \infty$. We
maintain $\gamma$ in the above expression, following the fact
that, experimentally, the reference $v_{\rm s}$ does {\sl not}
correspond to $\sigma_{xx}(q) =\infty$ (owing to impurities, and
$q\not\approx 0$).

In the present work we employ the `modified random phase
approximation' (MRPA) for $K_{00}(q,\omega)$ due to Simon and
Halperin \cite{Simon3} which takes account of the renormalisation
of the mass of CFs and which approaches the RPA for
$K_{00}(q,\omega)$ as $q\to 0$, the latter coinciding to leading
order (proportional to $q^2$) with the {\sl exact}
$K_{00}(q,\omega)$ for $q\to 0$ {\sl and} any $\omega$ at which
both $K_{00}^{\Sc r\Sc p\Sc a}(q,\omega)$ and $K_{00}(q,\omega)$
are bounded, which is the case for $\vert \omega\vert <
\Delta\omega_{\rm c} {:=} e \Delta B/m_{\rm b}$ (see in particular
Eq.~(5.6) in Ref.~\cite{Zhang0}). The $K_{00}^{\Sc m\Sc r\Sc p\Sc
a}$ has thus the property that it conforms with the requirement of
a Kohn theorem \cite{Kohn} according to which $K_{00}(q\to
0,\omega)$ must be determined by $m_{\rm b}$ rather than
$m_{\star}$. For the explicit expression concerning $K_{00}^{\Sc
m\Sc r\Sc p\Sc a}(q,\omega)$ corresponding to $\nu_{\rm e} = p/(2
n p + 1)$ in terms of elementary functions, we refer the reader to
Ref.~\cite{Simon3} (we have presented and employed $K_{00}^{\Sc
m\Sc r\Sc p\Sc a}(q,\omega=0)$ in Ref.~\cite{Farid1}). We mention
that for the purpose of calculating $K_{00}^{\Sc m\Sc r\Sc p\Sc
a}(q,\omega)$, which depends on $p$ and $n$, for a continuous
range of $\Delta B$ around zero, we first determine $\nu_{\rm e}$
from $\nu_{\rm e} = h n_{\rm e}/(e B)$, with $B$ the {\sl total}
applied magnetic flux density, and subsequently obtain the
required $p$ from $p = [\nu_{\rm e}/(1 - 2 \nu_{\rm e})]$ when
$\nu_{\rm e} < \frac{1}{2}$, and $p = [(1-\nu_{\rm e})/(1 - 2
(1-\nu_{\rm e}))]$ when $\nu_{\rm e} > \frac{1}{2}$, where $[x]$
denotes the greatest integer less than or equal to $x$. The
$\nu_{\rm e}$ corresponding to the thus-obtained $p$ remains
constant for certain ranges of values of $B$, which causes
artificial stepwise-constant behaviour in the functions of $\Delta
B$ that depend on $K_{00}^{\Sc m\Sc r\Sc p\Sc a}(q,\omega)$.

We model the effects of the impurity scattering through
substituting $\omega +i/\tau_0$ for $\omega$ in $K_{00}^{\Sc m\Sc
r\Sc p\Sc a} (q,\omega)$. Here $\tau_0$ stands for the scattering
time which is related to the mean-free path $\ell_0 = v_{\Sc f}
\tau_0$, where $v_{\Sc f}$ stands for the Fermi velocity. In
general, the substitution $\omega\rightharpoonup \omega +
i/\tau_0$ amounts to solving the Boltzmann equation within the
framework of the relaxation-time approximation, neglecting the
so-called current-conservation correction which has been found to
have {\sl no} significant consequences in contexts similar to that
of our present considerations (see Sec. 4.1.4 in
Ref.~\cite{Simon2}). With (from here onwards we suppress `{\Sc
m\Sc r\Sc p\Sc a}') $K'_{00}(q,\omega+i/\tau_0) \equiv {\rm
Re}\big\{K_{00}(q,\omega +i/\tau_0)\big\}$ and $\gamma' {:=} {\rm
Re}\{\gamma\}$, from the above-presented expression for $\Delta
v_{\rm s}/v_{\rm s} + i \kappa/q$ we obtain
\begin{equation}
\label{e3}
\frac{\Delta v_{\rm s}}{v_{\rm s}} = \frac{\alpha^2}{2}
\big( \gamma' - v(q) K_{00}'(q,v_{\rm s} q + i/\tau_0)\big).
\end{equation}
We eliminate $\gamma'$, whose value has {\sl no} influence on the
{\sl form} of $\Delta v_{\rm s}/v_{\rm s}$, by requiring that
$\Delta v_{\rm s}/v_{\rm s}$ coincide with the experimental
$\Delta v_{\rm s}/v_{\rm s}$ for $\Delta B\approx 0$.

{\it 3.~Computational results and discussions.---} In Fig.~1 we
present $\Delta v_{\rm s}/v_{\rm s}$ as a function of $\Delta B$
for the cases where $n_{\rm e} = 6.92\times 10^{10}$~cm$^{-2}$, $f
{:=} \omega/(2\pi) = 8.5$~GHz (left panel), to be compared with
Fig.~4 in Ref.~\cite{Willett2}, and $n_{\rm e} = 1.6\times
10^{11}$~cm$^{-2}$, $f = 10.7$~GHz (right panel), to be compared
with Fig.~1 in Ref.~\cite{Willett3}. The excellent agreement
between the theoretical results corresponding to an enhanced
$m_{\rm b}$ (with respect to $m_{\rm b}^0$) and experimental
results, in particular when these are compared with those obtained
within the same theoretical framework in which $m_{\star}$ retains
the same enhanced value as compared with $m_{\rm b}^0$ but $m_{\rm
b} =m_{\rm b}^0$ (curves (c)), strongly support the viewpoint that
under the conditions where $\nu_{\rm e} \approx \frac{1}{2}$, the
bare band mass $m_{\rm b}^0$ should be enhanced. This observation
is compatible with the experimental finding with regard to the
stronger than the theoretically-predicted divergence
\cite{Halperin} of the CF mass for $\nu_{\rm e}\to \frac{1}{2}$
\cite{Du2,Manoharan,Du3,Coleridge}. In this connection we should
emphasize that the closer one approaches $\nu_{\rm
e}=\frac{1}{2}$, the less sensitive $\Delta v_{\rm s}/v_{\rm s}$
becomes with respect to the further increase of $m_{\star}$ (or
$m_{\rm b}$ for that matter); our choices $m_{\rm b}=16 \times
m_{\rm b}^0$ and $m_{\star} = 0.5\times m_{\rm b}$ are based on
the consideration that the experimental features corresponding to
$\Delta B$ in the range $\sim 0.1 - 1$~T be reproduced. The
results in Fig.~1 obtained through employing the semi-classical
$\sigma_{\mu\nu}$, due to Cohen, Harrison and Harrison (CHH)
\cite{Cohen} (see also Appendix B in Ref.~\cite{Halperin} as well
as Eqs.~(2) and (3) in Ref.~\cite{Willett2}), show the inadequacy
of the semi-classical approach; curves marked by (d) unequivocally
demonstrate the shortcoming of strictly adhering to the viewpoint
that CFs behaved like non-interacting electrons exposed to a
reduced magnetic field --- curves marked by (b), which are
similarly based on the CHH $\sigma_{\mu\nu}$, owe their
resemblance to the experimental results to the fact that in their
calculation {\sl explicit} account has been taken of the
conditions which are specific to the regime corresponding to
$\nu_{\rm e}\approx \frac{1}{2}$ (see caption of Fig.~1). With
reference to our earlier work \cite{Farid1}, it is appropriate to
compare curve (a) in the right panel of Fig.~1 with the curves in
Fig.~1 of the work by Mirlin and W\"olfle \cite{Mirlin} and
compare both with the experimental trace in Fig.~1 of
Ref.~\cite{Willett3}. One observes that our present result, in
contrast with those in Ref.~\cite{Mirlin}, precisely reproduces
almost {\sl all} features of the experimental trace, such as the
values of $\Delta v_{\rm s}/v_{\rm s}$ at $\Delta B \approx 0,
0.38, 0.54, 1.0$~T.

From Eqs.~(\ref{e1}) and (\ref{e3}) we obtain
\begin{equation}
\label{e4}
\frac{\sigma_{xx}(q)}{\sigma_{\rm m}} = \Big(\big(\gamma'
- v(q) K'_{00}(q,v_{\rm s} q+i/\tau_0)\big)^{-1} -1\Big)^{1/2},
\end{equation}
according to which $\sigma_{xx}(q)$ interestingly does {\sl not}
explicitly depend on $\alpha^2/2$. Unless we set $\gamma'=1$, we
eliminate $\gamma'$ in Eq.~(\ref{e4}) by requiring that for given
values of $\sigma_{\rm m}$ and $q$, $\sigma_{xx}(q)$ according to
Eq.~(\ref{e4}) yield the corresponding experimental SAW velocity shift.
\begin{figure}
\includegraphics[width=3.4in]{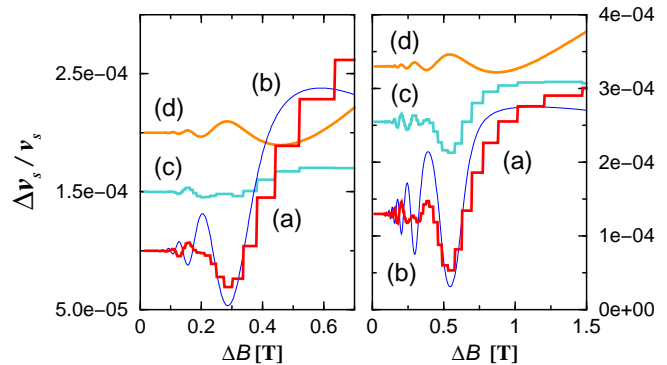}
\caption{\label{fi1}\rm (colour)
The relative shift of the SAW velocity. (a) and (c) are obtained from
Eq.~(\protect\ref{e3}), while (b) and (d) are based on
Eq.~(\protect\ref{e1}) in which the semi-classical result for
$\sigma_{\mu\nu}$, due to CHH \protect\cite{Cohen}, has been employed:
for (d), $\sigma_{xx}(q)$ has been directly used, whereas for (b)
$\sigma_{xx}(q)$ has been obtained from $\sigma_{yy}(q)$ according to
$\sigma_{xx}(q) \approx (\nu_{\rm e} e^2/h)^2\vert_{\nu_{\rm e}
=\frac{1}{2}}/\sigma_{yy}(q)$ --- this expression takes explicit account of
the fact that for $\nu_{\rm e} \approx \frac{1}{2}$, $\varrho_{xx}
\varrho_{yy} \ll \varrho_{xy}^2$ and $\varrho_{yy} \approx
1/\sigma_{yy}$; in both cases we have scaled $\sigma_{xx}/\sigma_{\rm m}$
by a constant such that for $\nu_{\rm e}=\frac{1}{2}$ the corresponding
$\Delta v_{\rm s}/v_{\rm s}$ coincide with the experimental value (for
clarity we have offset (d) by $10^{-4}$ in the left panel and by
$2\times 10^{-4}$ in the right). The equality of (a) and (c) (we have
offset (c) by $5\times 10^{-5}$ in the left panel and by $1.75\times
10^{-4}$ in the right) with the associated experimental results at
$\nu=\frac{1}{2}$ has been enforced through appropriate choices for
$\gamma'$ ($\approx 1.1$).
{\it Left panel}:
$n_{\rm e} = 6.92 \times 10^{10}$~cm$^{-2}$, $\ell_0 = 0.4$~$\mu$m,
$d=10$~nm, $f=8.5$~GHz.
{\it Right panel}:
$n_{\rm e} = 1.6 \times 10^{11}$~cm$^{-2}$, $\ell_0 = 0.6$~$\mu$m
(corresponding to $\tau_0 = 19.6$~ps),
$d=100$~nm, $f=10.7$~GHz.
{\it Both panels}:
(a), $m_{\rm b} = 16 \times m_{\rm b}^0$, $m_{\star} = 0.5
\times m_{\rm b}$; (c), $m_{\rm b} = m_{\rm b}^0$, $m_{\star} = 8
\times m_{\rm b}$. For the origin of the step-like behaviour in
curves (a) and (c) see the main text.}
\end{figure}
In Fig.~2 we present our theoretical $\sigma_{xx}(q)$ in
comparison with its SAW-derived experimental $\sigma_{xx}(q)$ by
Willett {\sl et al.} \cite{Willett1} (see the middle panel of
Fig.~2 herein). The details in Fig.~2 again support our above
finding with regard to $m_{\rm b}$ and $m_{\star}$, that a mere
enhancement of $m_{\star}$ with respect to $m_{\rm b}^0$ is {\sl
not} sufficient (see inset [B]). We also observe that a finite
$\ell_0$ is most crucial to the experimentally-observed linear
behaviour of the SAW-deduced $\sigma_{xx}(q)$ for $q$ in the range
$\sim (0.015/a_0, 0.075/a_0$), with $a_0$ the Bohr radius (see
curve (c)); the original observation with regard to
$\sigma_{xx}(q) \propto q$ for $q \gg 2/\ell_0$ \cite{Halperin}
thus turns out to be relevant for values of $q$ far outside the
experimental range. We note that $\ell_0 \approx 0.5$~$\mu$m,
employed by us, coincides with that reported in the pertinent
experimental papers. A further aspect that our present results in
Fig.~2 clarify is that, in contrast with earlier observations (see
the paragraph following Eq.~(7.6) in Ref.~\cite{Halperin} and that
following Eq.~(211) in Ref.~\cite{Simon2}), the available
experimental results by {\sl no} means are in conflict with the
predictions of Eq.~(\ref{e2}): our theoretical results for
$\sigma_{xx}(q)$ in Fig.~2 have been obtained through multiplying
$\sigma_{xx}(q)/\sigma_{\rm m}$ by {\sl the same} \cite{Willett4}
$\sigma_{\rm m}$ that has been employed to determine
$\sigma_{xx}(q)$ from the SAW-deduced $\sigma_{xx}(q)/\sigma_{\rm
m}$. Thus, rather than Eq.~(\ref{e2}) being inadequate, the
empirical method of determining $\sigma_{\rm m}$ (which employs
the {\sl dc} conductivity \cite{Willett3,Willett4}) should be
considered as inappropriate.

{\it 4.~Comment on the physical significance of the SAW
experiments.---} We now briefly focus on the physical significance
of the expression for $\Delta v_{\rm s}/v_{\rm s}$ in
Eq.~(\ref{e1}). To this end, let $\delta v_{\rm ext}({\pmb r},t)$
denote an applied time-dependent external potential, representing
that corresponding to SAWs. The change in the Hamiltonian of the
system, following the application of $\delta v_{\rm ext}$, has the
form $\delta \widehat{H}(t) = \int {\rm d}^2r\, \delta v_{\rm
ext}({\pmb r},t) \hat{\psi}^{\dag}({\pmb r}t) \hat{\psi}({\pmb
r}t)$, where $\hat{\psi}^{\dag}({\pmb r}t)$, $\hat{\psi}({\pmb
r}t)$ denote creation and annihilation field operators. Denoting
the change in the energy of the system corresponding to $\delta
\widehat{H}(t)$ by $\delta E(t)$, we have \cite{Farid2} $\delta
E(t) =$ $\int {\rm d}^2r$ $\delta v_{\rm ext}({\pmb r},t)
\bar{n}_e({\pmb r},t)$, where $\bar{n}_e({\pmb r},t) {:=} \int_0^1
{\rm d}\lambda\, n_{\rm e}^{(\lambda)}({\pmb r},t)$. Here $n_{\rm
e}^{(\lambda)}({\pmb r},t)$ stands for the instantaneous number
density of the system corresponding to $\delta v_{\rm
ext}^{(\lambda)}({\pmb r},t) {:=} \lambda \delta v_{\rm ext}({\pmb
r},t)$. By assuming $\delta v_{\rm ext}({\pmb r},t) \equiv \delta
v_{\rm ext}({\pmb r}) \exp(-i\omega_0 t)$, one obtains for $\delta
{\cal E}(\omega) {:=} \int {\rm d}t$ $\delta E(t) \exp(i\omega
t)$, $\delta {\cal E}(\omega) = -\int {\rm d}^2r {\rm d}^2r'\,
\delta v_{\rm ext}({\pmb r}) \overline{K}_{00}({\pmb r},{\pmb r}';
\omega-\omega_0) \delta v_{\rm ext}({\pmb r}')$, where
$\overline{K}_{00}({\pmb r}, {\pmb r}';\omega) {:=} \int_0^1 {\rm
d}\lambda\, (1-\lambda)\, K^{(\lambda)}({\pmb r},{\pmb
r}';\omega)$ \cite{Farid2}. Under the {\sl assumption} that
$\delta v_{\rm ext}({\pmb r})$ is weak, the dependence on
$\lambda$ of $K^{(\lambda)}_{00}$ can be neglected so that
$\overline{K}_{00} \approx \frac{1}{2} K^{(0)}_{00}$ where
$-K^{(0)}_{00}$ stands for the density-density response function
of the uniform, unperturbed, system. To second order in $\delta
v_{\rm ext}(q)$, one thus obtains $\delta {\cal E}(\omega) =
-\frac{1}{2} K_{00}(q,\omega-\omega_0) \vert \delta v_{\rm ext}(q)
\vert^2$. Let now $\overline{\delta E}_t {:=} \int_0^{t} {\rm
d}\tau\, \delta E(\tau)$, from which one readily obtains
$\overline{\delta E}_t \approx -\frac{1}{2} \vert \delta v_{\rm
ext}(q)\vert^2 \int {\rm d} \omega\, K_{00}(q,\omega-\omega_0)
\{1-\exp(-i\omega t)\}/(2\pi i\omega)$. For large $t$, the
integrand of the $\omega$ integral becomes highly oscillatory so
that to the leading order in $1/t$, $\int {\rm d}\omega\,
K_{00}(q,\omega-\omega_0) \{1-\exp(-i\omega t)\}/(2\pi i\omega)
\sim K_{00}(q,-\omega_0) \int {\rm d}\omega\, \{1-\exp(-i\omega
t)\}/(2\pi i\omega) = K_{00}(q,-\omega_0)$. Since
$K_{00}(q,\omega)$ is an even function of $\omega$, we eventually
obtain $\overline{\delta E}_t \sim -\frac{1}{2} K_{00}(q,\omega_0)
\vert\delta v_{\rm ext}(q)\vert^2$, for $t \gg 1/\omega_0$. This
expression, which coincides with Eq.~(14) in Ref.~\cite{Simon1},
is the fundamental link between $K_{00}(q,\omega)$ and $\Delta
v_{\rm s}/v_{\rm s} + i\kappa/q$ presented above. These details
make explicit firstly, that only small-amplitude perturbations are
correctly accounted for by Eq.~(\ref{e1}) (and similarly,
Eq.~(\ref{e3})), and secondly, that although, as suggested by
Willett {\sl et al.} \cite{Willett3}, the observation of
``geometric resonance'' and the ``cyclotron frequency deduced from
dc transport'' are inconsistent with ``a non-interacting,
semi-classical quasiparticle model [for CFs]'', they are in fact
{\sl not} inconsistent with the physical picture that the above
derivation brings out: that the mechanism underlying the SAW-based
experiments does not involve any resonance phenomenon in the usual
sense and that the SAW experiments, which involve a long-time {\sl
integration} of the fluctuations in the total energy of 2DESs,
unveil $K_{00}(q,\omega)$ at $\omega =\omega_0 \equiv v_{\rm s} q$
and $q=\omega_0/v_{\rm s}$, {\sl independent} of the magnitude of
the CF cyclotron frequency $\Delta \omega_{{\rm c}\star}$ and
consequently of that of the CF mass $m_{\star}$.
\begin{figure}
\includegraphics[width=3.4in]{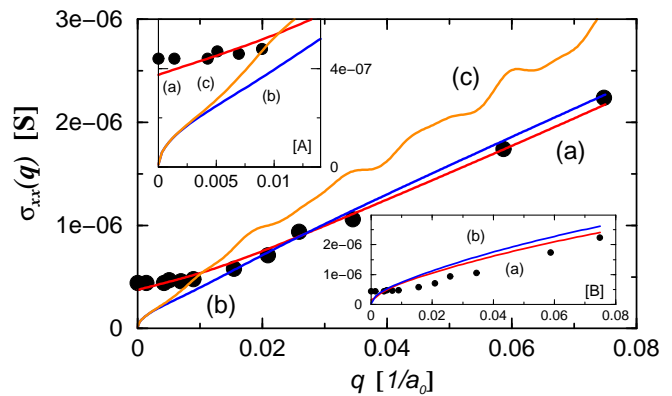}
\caption{\label{fi2}\rm (colour)
The longitudinal conductivity as deduced from the SAW
$\Delta v_{\rm s}/v_{\rm s}$ at $\nu=\frac{1}{2}$ (here $a_0 =
9.79$~nm) corresponding to $n_{\rm e} = 6.27 \times 10^{10}$~cm$^{-2}$
($k_{\Sc f}^{\Sc c\Sc f} = 0.87/a_0$). Solid dots are experimental
results by Willett {\sl et al.} \protect\cite{Willett1}, and (a),
(b) and (c) have been calculated according to Eq.~(\protect\ref{e4});
for all three cases we have assumed $m_{\rm b} = 16\times m_{\rm b}^0$,
$m_{\star} = 0.6 \times m_{\rm b}$, $d=10$~nm (for this value of $d$,
$\sigma_{\rm m} = 3.57 \times 10^{-7}$ ${=:} \sigma_{\rm m}^{\rm th}$
according to Eq.~(\protect\ref{e2})); in order to compare our results
with those in Ref.~\protect\cite{Willett1}, we have multiplied the
theoretical $\sigma_{xx}(q)/\sigma_{\rm m}$ by the value for
$\sigma_{\rm m}$ employed in this reference, namely
\protect\cite{Willett4} $\sigma_{\rm m} = 1.67 \times 10^{-6}$~S.
(a) has been determined with $\ell_0 = 0.6$~$\mu$m, (b) with $\ell_0
= 0.9$~$\mu$m and (c) with $\ell_0 = 2.4$~$\mu$m. (b) and (c) have been
calculated with $\gamma'=1$, while (a) has been obtained by following
the procedure outlined in the text: taking the experimental result
$\sigma_{xx}(q=0.004/a_0) = 0.44 \times 10^{-6}$~S, we have obtained and
used $\gamma'=0.952$. Inset [A] is a focus on the small-$q$ region of the
main diagram (the $q^{1/2}$-behaviour of (b) and (c) for $q\to 0$ is
associated with $v(q) K_{00}(q,\omega) \propto q$ in this region), while
inset [B] shows the counterparts of curves (a) and (b) for $m_{\rm b}
= m_{\rm b}^0$, $m_{\star} = 8 \times m_{\rm b}$ (for (a), $\gamma'
=1.008$). }
\end{figure}

{\it 5.~Concluding remarks.---} In conclusion, by strictly
adhering to the microscopic theory of CFs, we have established
that the SAW-based experimental results by Willett and co-workers
close to $\nu_{\rm e} =\frac{1}{2}$ can be remarkably accurately
reproduced provided the electron band mass $m_{\rm b}$ is
substantially enhanced with respect to $m_{\rm b}^0$; such an
enhancement is in conformity with the independent experimental
results by Shashkin {\sl et al.} \cite{Shashkin} concerning
two-dimensional electrons in silicon for $n_{\rm e}$ less than a
critical value approximately equal to $2 \times 10^{11}$
cm$^{-2}$. In this picture, the (observed) large value of the CF
mass $m_{\star}$ follows from a subsequent reduction of $m_{\rm
b}$ (owing to quantum fluctuations) rather than a direct
enhancement of $m_{\rm b}^0$. We have further established that a
finite mean-free path $\ell_0$ is essential to the
experimentally-observed linearity in the SAW-deduced
$\sigma_{xx}(q)$ in the range $1.56 \lesssim q \lesssim 7.68$
$\mu$m$^{-1}$, and that there exists {\sl no} discrepancy between
the theoretical and experimental values for $\sigma_{\rm m}$.

{\it Acknowledgements.---} I thank Professors D.~E.
Khmel'nitski\v{\i}, P.~B. Littlewood and H.~T.~C. Stoof for
discussion and Professor B.~I. Halperin and Drs S.~H. Simon and
R.~L. Willett for kindly clarifying some aspects concerning
$\sigma_{\rm m}$. \\

%



\end{document}